\documentclass[fleqn,twoside]{article}
\usepackage{espcrc2}

\usepackage{graphicx}
\usepackage[figuresright]{rotating}

\newcommand{\AmS}{{\protect\the\textfont2
  A\kern-.1667em\lower.5ex\hbox{M}\kern-.125emS}}

\title{The Solar--Stellar Connection}

\author{S. M. White\address{Department of Astronomy, University of Maryland, \\
College Park MD 20742 USA}}
       
\begin{document}

\begin{abstract}
Stars have proven to be surprisingly prolific radio sources and the
added sensitivity of the Square Kilometer Array will lead to advances in
many directions. This chapter discusses prospects for studying the physics of
stellar atmospheres and stellar winds across the HR diagram.
\vspace{1pc}
\end{abstract}

\maketitle

\section{Introduction}

The discovery that radio emission from stars is quite common and readily
detectable was one of the unexpected advances produced by the Very Large
Array \cite{Whi00b,Gud02}. This discovery could not be predicted based on what we
know of the Sun's radio emission, even though similar physical mechanisms
are operating in stellar radio sources 
and so knowledge of the Sun's emission is essential for
understanding stellar radio emission: this is the solar--stellar
connection in radio astronomy. Most of the obvious classes of star have been
surveyed, and many of them have been detected, including
the expected thermal wind
sources amongst both cool and hot mass--losing giant stars and 
symbiotic binaries, but also solar--like activity from 
magnetically active stars in the giant, subgiant, main--sequence
and (more recently) brown dwarf classes, as well as in chemically
peculiar B stars and pre--main--sequence stars. However, the total
number of stars detected is still relatively small, and selection
effects mean that in all cases we detect
only the most luminous tail of the most nearby members of the
populations of these objects,
representing a strongly biassed sample. For no populations of stars do
we have a complete radio sample, and particularly a completely detected
sample, that can be used to study the relationship of radio emission,
and by extension atmospheric structure, with other stellar properties such as
evolutionary state. 

Radio emission from stars comes from the atmosphere, and hence can be
used to study the nature of the atmosphere. Two
types of atmospheric emission are most common in the sources detected so
far: emission from an extended outflowing envelope or wind, or emission from
confined parts of the atmosphere such as the corona (where the
temperature exceeds 10$^6$ K) or chromosphere (the lower transition
region between the stellar photosphere and the hot corona, with
temperatures in the range 5--20 $\times$ 10$^3$ K).
Stellar outflows and chromospheres are generally detected via
the mechanism known variously as
{\it thermal free--free} or { \it bremsstrahlung} emission. This 
is responsible for the classic stellar--wind
radio emission. The opacity for this mechanism varies as 
$n_e^2\,T^{-1.5}\,\nu^{-2}$, where $n_e$ is the electron density, $T$
the electron temperature and $\nu$ the radio frequency.
Since the temperature of a stellar wind or a chromosphere 
is generally around 10$^4$ K,
large radio fluxes require very large optically thick surface areas: a 1
milliarcsecond disk, which is the order of magnitude for the 
photosphere of nearby giant stars, 
with a brightness temperature of 10$^4$ K
produces of order 1 microJy of radio flux at 10 GHz. For
the currently detected objects, the source size is much larger than 1
milliarcsecond due to high densities or very low outflow speeds for
the stellar winds, or very distended atmospheres in the case of
supergiants such as Betelgeuse \cite{LCW98,HBL01}.

{\it Nonthermal synchrotron emission} is the
basic microwave emission process operating in solar flares
and in stars which show solar--like magnetic activity.
This mechanism requires nonthermal distributions of
mildly (``gyrosynchrotron'') or highly
(``synchrotron'') relativistic electrons in magnetic fields of order
gauss or stronger. The brightness temperatures achieved by this
mechanism can be much larger than with bremsstrahlung, 
since the radiating electrons have much more energy than a thermal
electron in a 10$^4$ K wind.  With magnetic field strengths of hundreds 
of gauss typical of stellar coronae,
this mechanism can have a spectral peak (corresponding to the
transition from optically thick at low frequencies to optically thin at
high frequencies) in the microwave range, as observed for solar flares.
For much lower magnetic fields, such as one expects outside stellar
atmospheres, this mechanism should be optically thin at upper SKA frequencies
and the radio spectral index $-\alpha$ is then related to the nonthermal
electron energy spectral index $-\delta$ by the synchrotron relationship
$\alpha\,=\,(\delta\,-\,1)/2$.

At low frequencies {\it plasma emission} is also important: this is a
resonant process in which electrostatic Langmuir waves at the electron
plasma frequency, $\nu_p\,=\,9000\,\sqrt{n_e}$,
are driven to very high effective brightness
temperatures by coherent interaction with a beam or a loss--cone
pitch--angle distribution and then convert to propagating
transverse waves at $\nu_p$ or its harmonic, $2\nu_p$. Radiation at $\nu_p$ is
heavily damped by collisional opacity and so this emission is typically
seen only at low frequencies (i.e., low densities) 
or when the ambient plasma is very hot.

In this chapter, we will discuss likely advances in the field of stellar
radio astronomy to be expected from SKA. No attempt will be made at a
comprehensive review of previous results nor of priority: cited
references are given as examples in which more detailed histories of the
subject can be found. The chapter by G{\"u}del in the collection
``Science with the SKA'', edited by A. R. Taylor and R. Braun, covers
similar topics with a more organized review of previous results
than is presented here, including an excellent discussion of the radio
Hertzsprung--Russell diagram and SKA detection limits for stars.

\begin{figure*}[htb]
\includegraphics[]{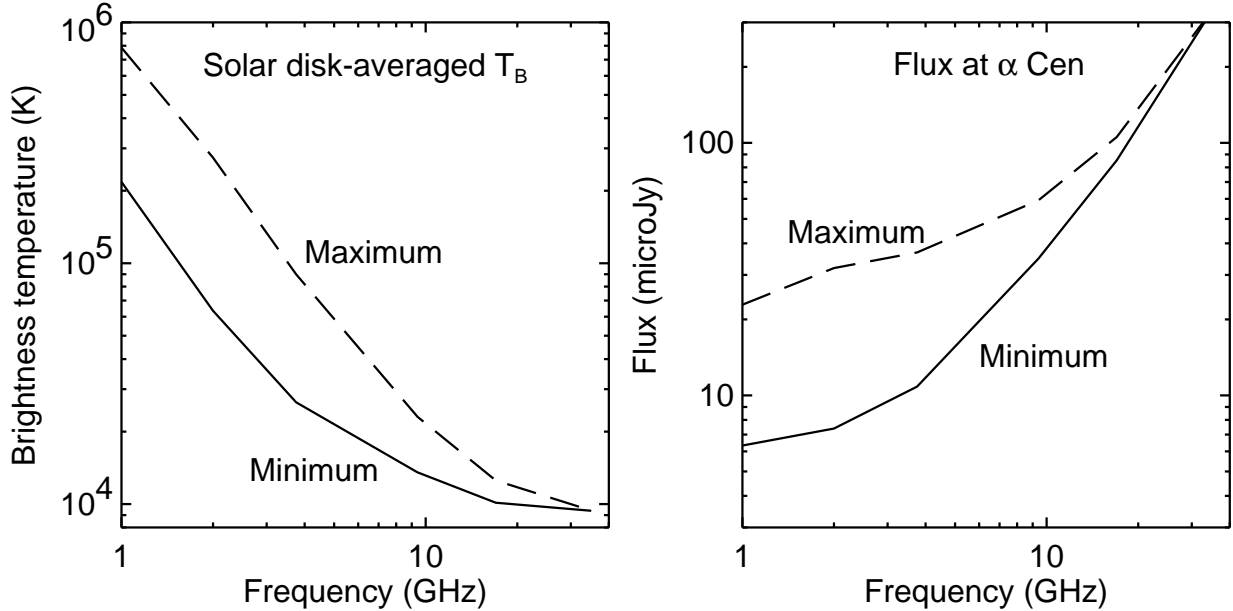}
\caption{The quiet Sun as a star at the distance of $\alpha$ Cen. The left
panel shows the disk--averaged brightness temperatures of the quiet
(non--flaring) Sun as a
function of frequency at solar minimum (solid line, May 1996) and solar
maximum (dashed line, December 2001) using fluxes measured by the
polarimeters of Nobeyama Radio Observatory. At right we show the same
fluxes moved to the distance of $\alpha$ Cen, 1.3 pc, in units of $\mu$Jy.
VLA detection limts for $\alpha$ Cen would be around 30 $\mu$Jy in 8
hours at 8 GHz 
if it were a northern--sky object, close to the predicted flux.
In 8 hours of observing SKA will have a detection limit well below 1
$\mu$Jy at 10 GHz: $\alpha$ Cen would be detectable out to 50 pc.}
\label{fig:stars}
\end{figure*}

\section{Main--sequence Stars}

At a distance of 1 AU, the quiet (non--flaring) 
Sun has a flux typically of order 3 $\times$
10$^6$ Jy at 10 GHz, with the flaring Sun rising to 10 times this value during
sporadic bursts that usually last tens of minutes or less. The
quiet--Sun flux at 10 GHz is equivalent 
to a disk--averaged brightness temperature of order 20000 K \cite{ZBH91},
corresponding to the solar atmosphere becoming optically thick at this
frequency in the chromosphere. At the
distance of $\alpha$ Centauri, the nearest star (1.3 pc) and a close solar
analog, these fluxes would be reduced to just 40
and 400 $\mu$Jy, respectively (Figure \ref{fig:stars}). 
These numbers illustrate the fact that
radio astronomy for truly solar--like stars is next to impossible with
current instrumentation, but SKA's sensitivity will open a much larger 
range of ``normal'' main sequence stars for study at radio wavelengths.
No main--sequence stars other
than the Sun have been detected so far through their thermal
atmospheres, although there have been specific searches 
\cite{WLK94,LWS96}, including $\alpha$ Cen; the only analogous
detection is of weak emission from the F5 subgiant Procyon that is 
attributed to its chromosphere \cite{DSB93}.

One of the important issues that a wider SKA survey will address is the
source of stellar coronal activity.
The dramatic result of VLA observations of cool
main--sequence stars is
that they possess something unknown on the Sun: a nonthermal corona
that produces strong and steady radio emission.  That is, in addition
to the thermal populations at 10$^{6-7}$ K in stellar coronae that
radiate
X--rays, there are nonthermal populations of electrons, extending up to
MeV energies, that are trapped on closed magnetic field lines and
produce strong radio emission. 
The Sun has no counterpart to this
population: such nonthermal electrons are usually
only present in the Sun's corona for very short periods during flares.
There is some evidence for small populations of lower--energy
long--lasting electrons
in the Sun's outer corona, producing ``storm'' emission at meter
wavelengths (probably via the plasma mechanism)
that is unrelated to the flare radio bursts \cite{KMS85}.
However, this seems to be very different from the process responsible
for the nonthermal radio coronae in active stars.

The nature of these nonthermal coronae and their
relationship to the thermal stellar atmosphere is poorly understood and
will be an important issue to be addressed by SKA.
The stars that are presently detected with this
emission are almost all young, rapidly--rotating stars active at other
wavelengths, such as in X--rays, and generally have magnetic
fluxes emerging through their photospheres that are many times the
magnetic flux in the solar photosphere. 
The SKA will allow us to detect both the thermal atmospheres and, where
present, the nonthermal atmospheres of a wide range
of main--sequence stars.  It should then be possible to
determine the dependence of the nonthermal atmosphere on stellar mass, age,
chemical composition and rotation rate. If we adopt a very conservative 
detection
threshold of 0.5 $\mu$Jy for SKA at 10 GHz (8 hour observation), then a
solar--radius--sized chromosphere with a brightness temperature of
10$^4$ K is detectable out to a distance of 50 pc: since almost all
stellar photospheres, except for those of white dwarves but including
brown dwarves, are about this large or bigger,
we can therefore expect a complete sample of detections of
thermal atmospheres out to this distance from SKA, encompassing many thousands
of stars. 

For the nearby sources with sizes larger than a few
milliarcseconds, including many red giants as well as a number of main
sequence stars, spatially resolved SKA images at the upper end of SKA's
frequency range, where fluxes are larger and the spatial resolution is
better, will add important additional information to the modelling.
Since only a few stellar atmospheres (really only Betelgeuse and Antares) can
be imaged by existing radio telescopes, SKA will greatly expand the
opportunities in this field.

In addition to the relationship between the nonthermal and thermal
components of stellar atmospheres, the SKA will yield spectra of the
thermal atmospheres that can be used to investigate the atmospheric
structure. In the case of the Sun, the radio spectrum of the quiet
atmosphere is dominated by the variation in temperature with height in
the solar atmosphere, since the height of the layer optically thick
in the radio decreases as the observing frequency increases.
The bremsstrahlung mechanism is well understood, and hence the radio
spectrum can be interpreted in terms of models of temperature and
electron density versus height. These data are crucial for modelling
stellar atmospheres because the radio data are taken in the
Rayleigh--Jeans limit, and therefore the observed brightness 
temperatures are true
temperatures. The radio data are used to constrain models based on UV line data
which are difficult to interpret on their own because the UV lines are not
formed in thermal equilibrium.
The standard solar atmospheric models have been
determined through a combination of UV and radio data
\cite{VAL76,GJL97}, and the same approach will also be required for stellar
modelling. With the large number of stars that SKA should detect,
atmospheric models can be constructed across a wide range of stellar types.

Monitoring of the stars within 50 pc will also likely lead to the
detection of stellar dynamo cycles in many main sequence stars. There is
somewhat of an anomaly in the present data: active stars detected at
X--rays do not show the order--of--magnitude modulation of the soft
X--ray flux from the corona that the Sun exhibits over its 11 year
cycle. However, since the stars we have good data for are mostly
very active stars, it is believed that their cycles are expressed
somewhat differently in the X--ray data. The quiet Sun also shows a clear
cycle in its radio flux. This is due to the presence at solar maximum of 
sunspot regions with dense hot plasma and strong magnetic fields that 
render the corona optically thick by bremsstrahlung at low frequencies
and by the gyroresonance mechanism (the non--relativistic
version of the gyrosynchrotron mechanism, operating typically at the
third harmonic of the electron gyrofrequency $\nu_B\,=\,2.8\times
10^6\,B_{\rm gauss}$ Hz) at higher microwave frequencies. 
This produces a significant area of optically thick emission above the
active regions,
at a brightness temperature of order 2 $\times$ 10$^6$ K,
in addition to the bremsstrahlung from the 
solar disk  at chromospheric temperatures (several
$\times$ 10$^4$ K). The solar--cycle modulation 
is more pronounced at lower microwave
frequencies (e.g., 3 GHz) because the
optically thick area at coronal temperatures is larger than at higher
frequencies (since the
required magnetic field strength or density is smaller) and the disk flux is
smaller: at 3.75 GHz the modulation is from 0.7 MJy at minimum to 2.0
MJy at maximum, while at 9.4 GHz it is from 2.6 MJy at minimum to 3.5
MJy at maximum (Fig.~\ref{fig:stars}). 
We can expect to detect such cycles on stars within 50
pc more easily than can be achieved by optical telescopes. 

Polarization
offers another valuable diagnostic in the radio regime: in the case of the Sun,
the active--region component of the emission from a given hemisphere 
is quite strongly circularly 
polarized due to the fact that the large leading spots
in active regions, where the strongest magnetic fields typically reside, almost
always have the same polarity in a given hemisphere over the 11 year
cycle, reversing in sense of polarization at the change of cycle.
While we likely will not be able to resolve the two hemispheres of cool dwarf stars with SKA, the
random orientations of stellar rotation axes means that in most cases we
will be viewing stars predominantly pole--on, and the emission will 
show the circular polarization appropriate to the hemisphere dominating
the emission. Any reversals due to dynamo cycles should then be
observable as reversals of the dominant polarization 
\cite{Whi02a}. The combination of these two techniques could
produce a much larger sample of dynamo measurements than we presently
possess.

\section{Young Stars}

Pre--main--sequence objects in star--forming regions are actually one of
the most common types of detected radio star: without having spent a lot of
time on them, we know of at least fifty detected stars in Taurus
\cite{CPL96,CMF97,MUM99} and several tens 
in Orion \cite{GMR87,FTC93}. Deep observations of star--forming
regions almost always produce a large number of faint continuum sources,
often cospatial with infra--red sources, that are almost certainly young
stars \cite{RAC99,StO98,RRA04}. Since most of the known regions of star
formation are over 100 pc away, they are difficult to study with the
VLA for reasons of sensitivity, but SKA will solve this problem and
provide a census of such sources. The field of view of SKA will make it
an efficient survey instrument for detecting embedded objects, and the
centimeter detections 
will provide an additional dimension to millimeter surveys of these regions with
ALMA, which will pick up primarily dusty disks.

Many of the stars detected in these regions show nonthermal radio
emission similar to that of nearby flare stars and active binaries.
However, there remain many other objects, particularly fainter sources, that do 
not fit this profile.
In Orion, several other classes of source have been identified
\cite{Gar87b}, including
partially--ionized globules that are associated with ``proplyds''
(proto--planetary disks), and
deeply embedded nonthermal objects. The sources of stellar outflows
often are very weak continuum sources, attributed to gas partially ionized
by shocks in the outflows \cite{Ang95,RRA04}: see also 
Hoare (this volume).  Classical T Tauri stars show
weak microwave emission, but it has quite different characteristics from
the nonthermal emission exhibited by the weak--lined T Tauri stars
that dominate radio surveys of star--forming regions: they are 
believed to be coeval with classical T Tauris but no longer in
possession of disks.  The conventional explanation
for the nonthermal emission in these young sources is magnetic activity driven
by the rapid rotation of a recently contracted and highly
convective star: while it is present, a protostellar disk controls and
limits the spin of the star, which can increase rapidly once the disk
is gone.

While we have detected many sources in star--forming regions, they are
mostly very faint detections and this limits our understanding of the
sources. Frequently they are only detected in the 5--8 GHz range and
spectral information is then limited. SKA will completely change this
situation, producing a complete census for many star--forming regions
and allowing a greater understanding of the nature of these sources and
investigation of the underlying physics.

\section{Stellar Flares}

In most respects, stellar radio flares so far detected 
from solar--like stars have resembled phenomena that we recognize from
the Sun:

\begin{itemize}

\item At lower microwave frequencies we tend to see highly ($\sim$100\%) 
circularly polarized flares with inferred brightness temperatures (assuming source
sizes comparable to the stellar disk) in excess of 10$^{10}$ K
\cite{SFG85,LMP88,OoB94}. Such high
brightness temperatures in combination with high polarization 
are not consistent with an incoherent
mechanism such as synchrotron emission, so they are almost certainly due
to a coherent process, and more likely plasma emission than electron
cyclotron maser emission because the polarization of the low frequency
flaring is opposite to that of the high frequency flaring. This is
expected if the low frequency flaring is plasma emission but the high
frequency flaring is synchrotron emission \cite{WhF95}. As is also true
on the Sun, this coherent low--frequency emission need not be related to
flares at other wavelengths: it represents electrons accelerated in the
stellar corona by some unknown form of energy release that does not
show up at optical or X--ray wavelengths.

\item At higher microwave frequencies we see analogs of solar microwave
bursts, mildly circularly polarized with a spectrum peaked at centimeter
wavelengths and extending as a power law to higher frequencies,
consistent with synchrotron emission from electrons
of energies from hundreds of keV upwards. On the Sun, the microwave
bursts are invariably associated with solar flares and X--ray bursts,
with the implication that all the observed high--energy phenomena are
due to electrons accelerated in the flare energy release.
In active stars, the correlation between flaring at different
wavelengths is not as strong: when the X--ray flare
rate is high, so is the radio flare rate \cite{BPB03,FSN03}, and sometimes radio
flares are seen in conjunction with X--ray flares, but frequently 
flares in both wavelength ranges occur within a given period but 
with no obvious connection linking them \cite{GBS96,OBA04}.

\item VLBI observations have revealed another difference between solar
radio bursts and stellar radio flares: the source sizes of stellar radio
bursts grow with time
and can reach values much larger than the stellar disk \cite{TUM93}, 
whereas microwave flares
on the Sun never achieve sizes significantly greater than the active
region in which they occur. This suggests a mechanism in which the
energy density of nonthermal particles in the radio source is so high
that it behaves like a plasmoid (a self--contained volume of magnetic
field entraining nonthermal particles) that evolves somewhat like a
fireball.

\end{itemize}

There is one important problem with the ``solar'' interpretation of the
higher-frequency stellar microwave flares given above: while the low--frequency
spectrum is rising as expected for an optically thick synchrotron source, 
the high--frequency spectra of these events tend to be embarrassingly
flat during the decay phase 
and are not consistent with the spectra expected of nonthermal
distributions of electrons accelerated in flares. The spectrum of the
quiescent (non--flaring) emission from active stars, also attributed to
synchrotron emission, also shows these flat high--frequency spectra.
The resolution of this problem is not clear. Some solar radio bursts
have spectra that rise from low frequencies well into the millimeter
domain \cite{KCC85}, but these events are exceedingly rare.

SKA will clearly identify radio flare phenomena, presently unknown due to
sensitivity limitations, across wide ranges of stellar types and allow
comparison with solar--flare activity. SKA will be able to follow
variations in flares on timescales an order of magnitude shorter than is
presently possible, providing a much clearer picture of the underlying
physics. This will permit us to understand
the nature of energy releases in the atmospheres of stars of different
masses and ages, and provide another important piece of the puzzle that
is stellar behaviour. SKA will also have the sensitivity to make dynamic
spectra of stellar flares, i.e., spectra as a function of time, that can
reveal electron beams or other disturbances such as shocks
propagating through the stellar coronae.

\section{Stellar Winds}

This topic is also covered in the chapters by Hoare and Marvel (this
volume). Stellar
winds come in many different guises. In hot stars, powerful line--driven
winds carry a large momentum and 
can have a major effect on the stellar environment, blowing
bubbles in the local medium and creating dense shells. 
Cool supergiants such as Betelgeuse have very slow outflows, thanks to
the very low gravity (and hence escape velocity) at their distended 
photospheres. Main sequence stars like the Sun have relatively 
low--momentum winds, but with quite high velocities ($\sim$ 500 km/s).
These winds are described as ``coronal'' winds because they are not
radiatively driven like the winds of very luminous stars, but rather
exist as a result of the pressure in the stellar atmosphere via Parker's
mechanism. 

Very little is actually known about the winds of other cool main sequence
stars because they are weak and very difficult to measure with existing
techniques: the best data come from absorption in 
the wings of the Lyman $\alpha$ line observed with HST \cite{WLM01,WMZ02}, 
and they show
winds with properties very similar to the Sun's 
(2$\,\times\,$10$^{-14}$ M$_\odot$ yr$^{-1}$) in $\alpha$ Cen (G2V)
but much weaker in Proxima Cen (M5.5Ve). The radio emission from
solar--like winds is weak.  For a spherically symmetric,
constant--velocity, fully--ionized coronal wind, we find 
\begin{eqnarray*}
\lefteqn{S \approx 1\ \mu{\rm Jy}~({\nu \over 10 {\rm \ GHz}})^{0.6}~
({T \over 10^6 {\rm \ K}})^{0.1}~} \\
\lefteqn{\ \ \ \ \ \ ({\dot M \over 10^{-13}~M_{\odot} {\rm \ yr^{-1}}})^{4/3} 
({v_\infty \over 300 {\rm \ km \ s^{-1}}})^{-4/3}~d_{pc}^{-2} }
\end{eqnarray*}
\noindent where $v_\infty$ is the terminal velocity of the wind, $T$ the
temperature of the wind, and $d_{pc}$ the distance to the star in
parsecs \cite{PaF75,WrB75}. This suggests that winds an order of
magnitude more powerful than the Sun's would be detectable by SKA at
least for some nearby stars,
opening the possibility of improving our knowledge of 
mass loss rates for cool main sequence stars. However, there is a simple
argument that states that for any star in which we can see coronal radio
emission, we cannot detect the wind at radio wavelengths 
because the wind detection relies on an optically thick surface in the
wind. However, if we see coronal emission, then the wind must be optically
thin \cite{LiW95}.
Since so many nearby stars are detected via their coronal or
chromospheric emission, this argues that essentially none of the cool
low--luminosity stars have mass loss rates much exceeding the Sun's.

Two very different emission mechanisms are responsible for radio
emission from hot star winds.
Thermal free--free is responsible for the classic stellar--wind
radio emission.  As noted earlier, the opacity for this mechanism varies as 
$n_e^2\,T^{-1.5}\,\nu^{-2}$.
The fact that the opacity decreases as frequency increases, while density
in an outflow decreases with radius, leads to a fundamental property of
free--free radio emission from stellar winds: higher frequencies probe
deeper into the stellar wind. For a constant--velocity wind
($n_e \propto r^{-2}$) the radius of the optically thick
surface, which limits how deeply we can see,
scales with frequency as $\nu^{-0.7}$.  The combination of this scaling
of optically--thick source dimension with frequency and the
$\nu^{+2}$--dependence
of the black--body emission law produces the classic $\nu^{0.6}$
spectrum of a constant--velocity free--free--emitting stellar wind.

In addition to the thermal wind emission, many sources also show
nonthermal synchrotron emission.  It is important in hot stars because
of the facility with which shocks can form  in the powerful winds:
the characteristic speed of a
hot star wind is generally in excess of the ambient sound speed in the
wind, so that any significant velocity fluctuation, such as a fast knot
overtaking a slower one or two winds colliding,
has the potential to result in a
shock \cite{LuW80}. Once a shock forms, electron acceleration apparently takes place
(by an as yet not completely understood mechanism). A high Mach--number
shock produces a characteristic power--law energy distribution of spectral
index -2, which results in a $\nu^{-0.5}$ radio flux spectrum (in the
optically--thin limit believed appropriate to these sources).
A detailed model for nonthermal emission from a single--star wind
carrying random shocks has been developed \cite{Whi85}.
Note that the acceleration must take place at some
distance from the star: the powerful
radiation field of a hot star can quench shock acceleration close to the
star where the inverse--Compton mechanism depletes energy from a
high--energy electron faster than a shock can supply it \cite{ChW94}.
This is a mild problem for models in which the
magnetic field in the synchrotron source
is a stellar field carried out by the wind, since it will
diminish with distance from the star. An alternative is for the magnetic
field also to be generated in the shocks as a byproduct of the plasma
turbulence there. There is currently a healthy debate as to whether
nonthermal radio emission is only seen in binary systems \cite{DoW00}, 
implying that
it may be due to the colliding--winds scenario and that the single--wind
model is ineffective.

Clearly SKA will detect many more winds from luminous stars than we
presently know of and greatly expand the range of stellar types for
which mass loss rates are known: for luminous stars the wind flux is 
\begin{eqnarray*}
\lefteqn{S \approx 70\ \mu{\rm Jy}~({\nu \over 10 {\rm \ GHz}})^{0.6}~
({T \over 10^4 {\rm \ K}})^{0.1}~} \\
\lefteqn{\ \ \ \ \ \ ({\dot M \over 10^{-7}~M_{\odot} {\rm \ yr^{-1}}})^{4/3} 
({v_\infty \over 300 {\rm \ km \ s^{-1}}})^{-4/3}~d_{kpc}^{-2} }
\end{eqnarray*}
\noindent where $d_{kpc}$ is the distance to the star in kiloparsecs. 
However, in this field, most work will involve collaboration with ALMA
because the combined frequency range of the two instruments probes a
wide range of radii within the wind and will be needed to investigate
effects such as radius--dependent variations in the ionization fraction
that are important in very dense winds. Observations of the spectrum are
also needed to separate out the relative contributions of nonthermal
emission with a falling spectrum and thermal wind emission with a rising
spectrum. 

Spatial resolution will also be very important for the study of stellar
winds. The famous mass--losing B[e] star MWC 349
is an example of the surprises in store when we have sufficient spatial
resolution: this star was the prototypical thermal  stellar wind radio 
source whose $\nu^{0.6}$ spectrum was for many years regarded as
perfectly explained by spherical stellar wind 
models. However, when the VLA finally resolved the star, it was found to 
be anything but spherical, showing at high resolution a curious
edge--brightened polar outflow apparently associated with a dust disk
\cite{DrW83}.
Another example is the Wolf--Rayet binary system WR 140 (HD 193793),
which shows both thermal and nonthermal components in the
system simultaneously. However, 
with an orbit dimension of only 26 mas the system is too small for the
VLA to resolve and despite being one of the brightest WR systems in the
radio, it is barely strong enough to be observed successfully on long
baselines with MERLIN or VLBI.
The improvements in both sensitivity and spatial resolution 
afforded by SKA will allow us to
study the spatial structure of the outflows of many more systems, both
single and binary, and to understand their properties in a systematic
fashion impossible with the few examples that we presently have.  

This work was supported by NSF grants AST 01--38317, ATM 02--33907 and ATM 03--20967, 
and NASA grants NAG 5--12860 and 5--12732.


\end{document}